# Gender neutrality in robots:
# An open living review framework

**KATIE SEABORN**

*Industrial Engineering & Economics*
*Tokyo Institute of Technology*

**PETER PENNEFATHER**

*gDial, Inc.*

# Gender Neutrality in Robots: An Open Living Review Framework

Katie Seaborn
*Department of Industrial Engineering and Economics*
*Tokyo Institute of Technology*
Tokyo, Japan
0000-0002-7812-9096

Peter Pennefather
*gDial, Inc.*
Toronto, Canada
0000-0002-2953-8977

*Abstract*—Gender is a primary characteristic by which people organize themselves. Previous research has shown that people tend to unknowingly ascribe gender to robots based on features of their embodiment. Yet, robots are not necessarily ascribed the same, or any, gender by different people. Indeed, robots may be ascribed non-human genders or used as "genderless" alternatives. This underlies the notion of gender neutrality in robots: neither masculine nor feminine but somewhere in between or even beyond gender. Responding to calls for gender as a locus of study within robotics, we offer a framework for conducting an open living review to be updated periodically as work emerges. Significantly, we provide an open, formalized submission process and open access dataset of research on gender neutrality in robots. This novel and timely approach to consensus-building is expected to pave the way for similar endeavours on other key topics within human-robot interaction research.

*Keywords*—robots, gender, robot gender, gender neutrality, genderless robots, mechanical genders, living review, open data

## I. Introduction

Gender is an important feature of people—and increasingly of robots, too. Decades of work has shown that people can and do read human characteristics in the design of robots and other agents, even when anthropomorphic cues are subtle. Nass, Brave, Moon, Lee, and colleagues [1], [2] were forerunners in researching this phenomenon. Their work on computer voice led to the Computers Are Social Actors (CASA) model, a widely recognized descriptive paradigm that shows how people tend to react to computer agents that have humanlike characteristics as if they are people, usually unthinkingly and in stereotyped ways. The implications of this early research were vast, spurring studies on gender in robots and other artificial agents [3]–[6].

While most researchers have taken a binary approach to gender, i.e., male/female or masculine/feminine or man/woman, an increasing number are calling for and starting to conduct research on alternative models of gender, including ambiguous, non-binary, fluid, mechanical/robotic, and neutral genders [7]–[9]. The notion of gender neutrality in robots has been raised for its potential to disrupt negative gendered associations [9] and model new ideas of gender, especially for non-human agents [7]. Others have argued that robots, especially humanoid and social robots, cannot be genderless due to the CASA phenomenon: our tendency to gender robots when humanlike cues are present [8]. Aldebaran-SoftBank's Pepper robot is a case in point. Although the designers aimed to create a gender-neutral robot in voice and body [10], the official materials clearly ascribe a masculine gender to Pepper by using he/his/him pronouns [1]. These complexities, tensions, contradictions, and possibilities have spurred interest on gender neutrality in robots. Stanford's *Gendered Innovations* group, for instance, includes "genderless robots" as a key consideration for roboticists and researchers who wish to explore gender in robots [6]. Gender neutrality is expected to be a central topic on robot gender going forward.

As work on this topic emerges, building consensus and identifying points for future work will be essential. The standard way to do this is by conducting a *systematic review* of the literature [11]. Systematic reviews are considered the gold standard of evidence in many scientific fields because they do not rely on a single study, instead bringing together all of the evidence in a standard, structured, and rigorous way [12]. Given the evolving nature of gender neutrality in robots as a topic of study, a type of systematic survey called a *living review* may be particularly appropriate. Living reviews are "alive" in the sense that they are periodically updated with new articles, findings, and possibly reporting structures [11], [13]. Living reviews are especially appropriate for cutting-edge topics where there is uncertainty in the corpus of evidence and a need for direction and congruency. This makes the topic of gender neutrality an ideal candidate for a living review. Moreover, living reviews can be an open science initiative if the procedures are registered and the datasets are provided in an *open access* venue. Doing so increases transparency and integrity, allows for the latest articles to be indexed in a structured way, and creates opportunities for community engagement, such as crowdsourcing new articles for inclusion. To the best of our knowledge, there is no living review work on the topic of gender neutrality in robots, or indeed in HRI generally, as well as no open access dataset based on living review work and/or community contributions.

As an open science initiative, we propose an open living review framework for generating and regularly updating an open access dataset of research on gender neutrality in HRI. We offer a structure for reporting on data relevant to gender neutrality: demographics, theory, methodology, results, and reflexivity. We provide a formalized means of submitting articles to the dataset. Our contributions are threefold: (1) a rigorous approach based on an international standard for continued and sustainable consensus-building on a topic of emerging significance to the HRI community; (2) the means of carrying out this process, including a formalized submission process; and (3) an open access dataset with initial data generated by following this approach. We expect this framework, along with the facilitating tools and open dataset, to be updated with community involvement. If successful and useful, it may be transferable to other topics of interest within the HRI community.

---

[1] https://www.softbankrobotics.com/emea/en/pepper

## II. CONCEPTUALIZING GENDER NEUTRALITY

Gender is a common yet complicated subject. *Gender* refers to a multidimensional set of features around which people categorize themselves and organize society [14]–[16]. Many models exist, both academically and colloquially. Typically, gender is distinguished from *sex*, with gender used to describe sociocultural qualities and constructs, such as identity and roles within society, and sex referring to aspects of biology and physiology [15], [16]. Research across fields of science has suggested that sex and gender can overlap [16]. Many societies rely on a *binary model* of gender/sex, premised on male/masculine/man and female/feminine/woman poles. Yet, a range of intersexes and genders within and beyond the binary exist, including transgenders, non-binary genders, gender fluidity, third genders, and more [16], [17]. Gender and sex often align for many people, i.e., people with biological characteristics that fall into the male category also tend to have a male gender identity. As a *social characteristic*, gender is also a matter of perception, based on mental models within individuals and societies. Indeed, this seems to drive phenomena such as the CASA when it comes to non-human but humanlike agents, including robots. In addition, because of its sociocultural origins, gendering can induce biased responses that are independent of biology [14]. This is especially true when robots are serving in roles that have been traditionally gendered.

*Gender neutrality*, at its most basic, refers to the absence of gender/ing [18]. In this sense, it is *relational*: it exists with respect to at least one other alignment. In the binary model of gender, these alignments are female/feminine/woman and male/masculine/man. *Gender neutral* can refer to an identity that one has or a characteristic that is ascribed by others. As an umbrella term, it may be used to encapsulate a range of gender alternatives: a lack of gender/ing, an avoidance of gender/ing, negating or downplaying gender/ing, allowing for gender to be ascribed in diverse ways, among others. Gender neutral language, for instance, removes all traces of gendering. Compare "meteorologist" to the male-gendered "weatherman": the former is genderless because no linguistic gender markers, such as "man," are present. At a high level, gender neutrality displaces the notion of gender and the act of gendering.

Gender neutrality in robots remains an open question. Roboticists and researchers may assume that robots are gender neutral by default. However, research has shown that robots can signal gender and/or elicit gendering through *cues* in their morphology, notably body [19] and voice [8], and in the interaction context, typically due to gender stereotyped views about the role or activity that the robot is carrying out [4], [20]. Given our propensity to gender, especially in accordance with the dominant gender binary model, it is difficult to define what constitutes gender neutral cues or even an absence of gender cues. Moreover, gender cues may not be perceived in the same way from person to person or culture to culture. Robertson, for instance, found that the Wakamaru robot was perceived to be feminine by Westerners and masculine by Japanese people, who viewed its form factor according to cultural models of dress [21]. As Sutton argues for voice assistants [8], it may be better to view robots s gender *ambiguous* until they are ascribed gender, and then consider what factors go into how these ascriptions are made and what the results are, especially any negative ones.

Other genders and ways of gendering may be considered within the gender neutral space. *Gender fluidity* refers to a change in gender identity or expression over time [17]. In their case study on social robots, Stanford's *Gendered Innovations* group suggests that robots could be designed in a gender fluid way, especially to mitigate negative gender stereotypes [6]. *Androgyny* refers to a mixture of masculine and feminine traits within one individual, or one robot in this case. Androgyny may create a balance or negating effect when gender is viewed from a binary perspective. *Agender* and *gender-free* are human identity complements to genderless robots. *Non-binary*, *third genders*, and *genderqueer* refer to gender identities that are not premised on the gender binary. For robots, the complements of these may be *mechanical* or *robotic* genders [7], in theory. These may not apply if people ascribe a range of mechanical genders to robots, which is not yet known.

At present, little is known about how gender neutrality has been approached in the design of robots and human-robot interaction (HRI) research. Even so, the work so far points to both diversity and convergence. The next step is to map out the state of affairs in an open and accessible way.

## III. DATASET

The dataset we are contributing is an open access repository of research on the topic of gender neutrality in HRI. This dataset will be regularly updated as part of a living review process. As of this writing, the first version of the dataset, generated by a rapid review, is provided. Details of the rapid review will be discussed later in the paper as a case study in performing a living review cycle. Data analysis will be reported in another paper.

We chose the Open Science Framework (OSF), created by the Center for Open Science (COS)[2], to host our dataset. Our decision was based on Nature's recommendations for data repositories[3]. OSF is a researcher-supported, open access data repository that offers version control, persistent storage, and permanent identifiers. There are few restrictions on the data, and it allows the data providers to choose a license. It also provides a means of sharing the data anonymously for peer review.

The dataset files and structure are as follows:

### A. Screening of Abstracts

This spreadsheet represents the first phase of a review cycle. It includes all results returned from database searches, alternative search tools (such as Google Scholar), manual additions, and, in our case, submissions from the community (see IV.A.1). It does not include duplicates or inappropriate types of records, which should be eliminated first. It includes the article title, authors, year of publication, abstract, decision on inclusion based on the abstract, reason for exclusion, which records were checked by a second reviewer (ideally 20% or more of the records excluded by the first reviewer), reason for disagreement between reviewers, and relevance despite exclusion (such as theoretical papers that may be cited).

---

[2] https://www.cos.io

[3] https://www.nature.com/sdata/policies/repositories#general

### 1) Screening of Full Text

This spreadsheet represents the second phase of a review cycle, wherein the full text of the article is reviewed based on the eligibility criteria (detailed below). The included or uncertain records from the previous phase, particularly the author, title, year of publication, and abstract details, are included. Additionally, inclusions at this stage, reason for inclusion or exclusion, new manual additions based on citations or references, and relevance despite exclusion are included.

### 2) Data Extraction

This spreadsheet represents the final included articles. An ID is assigned to all articles that are brough forward from the previous stage. Author, title, and year of publication are included, as well as all data relevant to the topic of gender neutrality in HRI research. These data were determined by two researchers with experience in HRI research and literature review methodology. They include: demographics (total sample size, total men, total women, totals for other genders, totals for unstated/unreported, age mean, age range); robot information (robot/platform name/s, voice origin, simulation or not); theory (theory name/s, previous work, brief description, citation/s); methods (manipulation check, other methods for establishing gender neutrality, gender options for participants, gender options for the robot/s, assessment of robot body, assessment of robot voice); results (in support of gender neutrality, against gender neutrality, other relevant results, relationship between robot voice and body); reflexivity (researchers' reasons for exploring gender neutrality, researcher's ascriptions of robot gender, participants' ascriptions of robot gender, gender neutral design cues, differences between researcher and participant ascriptions of robot gender); and any additional notes.

### 3) Risk of Bias / Quality Assessment

This spreadsheet includes the results of a risk of bias or quality assessment tool for the included articles. All items from the instrument as well as a final score or decision are included next to the article's author, title, and publication year.

### 4) Accessing the Dataset

The dataset can be accessed here: https://osf.io/v6fwg/

## IV. LIVING REVIEW FRAMEWORK AND METHODOLOGY

We have adapted the Cochrane living review guidelines as a basis for our living review framework and methodology. First, we will present our framework, which comprises the steps for carrying out the living review and generating the dataset. Then, we will briefly outline the methodological details that we have customized for the topic of gender neutrality in HRI research.

### A. Living Review Framework

Our living review framework comprises the following steps:

#### 1) Step 1: Data Collection

The first step is data collection, i.e., gathering articles, using a combination of traditional systematic search methods (i.e., searching databases with keywords and queries) and community-driven crowdsourcing of articles through a Google Form. This form asks a series of questions about the article that are directly tied to the columns in the dataset. Contributors are asked to fill out as much detail as possible to ease the screening and data extraction process. The submission form is available here: https://forms.gle/bipvG34mhU2hzhiEA

#### 2) Step 2: Screening the Data

The next step is to screen the data for eligibility, including checking for duplicates. This is described in more detail below.

#### 3) Step 3: Updating the Dataset

After the article has been screened and deemed eligible, the provided data will be checked. Any blanks will be filled in, where possible, and any incongruencies will be corrected. The dataset will then be updated on OSF.

#### 4) Step 4: Periodic Publication of Updates

After a significant number of updates to the dataset, estimated to be within a 2-year span, we will integrate the new data into existing analysis and/or generate new analyses. These will be peer-reviewed and published in an appropriate venue.

### B. Living Review Methodology

All Cochrane reviews, including living reviews, require the following components and steps to be taken and reported on. These components are presented in line with PRISMA [22]. We describe each in turn and offer details specific to the topic of gender neutrality in HRI research. In essence, this is a customized methodology fit for our purposes; readers are encouraged to refer to the Cochrane materials and/or Elliott et al. [11] for full details on the basic living review methodology.

#### 1) Eligibility Criteria

Papers are included if they report on human subjects research in HRI where gender neutral voice was considered in the evaluation of robots. Papers also need to be published as a full or short study in a peer-reviewed venue. Papers are excluded for these reasons: No human subjects findings on gender neutral voice reported; gray literature, unpublished reports, preprints, non-peer reviewed papers, conference papers and talks without proceedings; not in the languages known by the authors.

#### 2) Information Sources

Information sources are typically major academic databases but can also include alternatives (such as Google Scholar), registries, and experts (who recommend articles for manual inclusion). All information sources should be listed. We will use, at minimum, academic databases representing general and engineering topics: Scopus, Web of Science, IEEE Xplore, and ACM Digital Library. Search dates need to be reported. Bibliographies of included papers should also be searched.

#### 3) Search Terms

Search terms are the keywords that make up the queries used in searches. They should be focused but comprehensives. For this topic, we use the following search terms: robot* and gender neutral* or neutral gender* or gender-neutral or genderless or gender-less or without gender or no gender* or gender ambigu* or mechanical gender* or agender or androgynous. The full queries used for each information source should be reported in an appendix. One researcher can run all queries.

*4) Study Selection*

Selection of articles involves a two-phase screening process with at least two reviewers. First, abstracts of potential papers are dual-screened independently by two or more reviewers. The primary reviewer should screen all records. About 20% of the same records or records excluded by this reviewer should be assessed by a second reviewer. Conflicts are resolved by discussion as they arise. Then, the full text of each paper is independently assessed for inclusion by the first reviewer based on the eligibility criteria. The second reviewer double-checks excluded papers. Disagreements are resolved by consensus.

*5) Data Extraction*

At least two authors extract data for an even portion of the papers. Each checks the extractions of the other.

*6) Risk of Bias and Quality Assessment*

Given the diverse study designs in HRI research, we will use the 13-item Quality Assessment for Diverse Studies (QuADS) tool [23]. Each reviewer independently evaluates the quality of articles included at the full text stage using a 3-point scale. Severe disagreements are discussed until consensus is reached.

*7) Data Analysis/Synthesis*

Data analysis methods should be reported, including any meta-analyses. Changed or new results should be positioned against the results reported in previous cycles.

*8) Dataset Structure Updates*

Living reviews bring in new work, which may include new factors to be recorded in the dataset. However, the dataset may not provide the structure for recording these new factors. That cycle of the living review should report on any changes to the dataset and ensure that the Google Form is updated as well.

*C. Governance and Transparency*

A living review must live on. We, the authors, should not be the sole arbiters; it should be a community effort with shared responsibility. Conflicts of interest may also arise; we should not necessarily review our own work. We encourage everyone to view this initiative as an essential effort for the good of the HRI community and knowledge production. Research may be carried out study by study, but the gold standard for determining the current state of general knowledge is the meta-analysis/synthesis of these individual studies as generated by rigorous review work. We should not shy away from critical opinions, outdated findings, or negative results. This is our duty as researchers.

Gathering reviewers will be a multi-pronged, ongoing effort. Workshops at HRI venues (e.g., RO-MAN, which has a gender workshop[4]) can "filter in" experts who have a vested interest in the topic via self-selection. Reviewer criteria will be based on a modified form of the HRI2022's reviewer criteria to start, and be refined later with community involvement: Must have worked in HRI for 2 or more years; or must have published at a peer-reviewed HRI venue; or must have worked in gender studies for 2 or more years; or must have published on gender as a focus of study in a peer-reviewed venue; ideally has experience doing review work. Those interested in being reviewers can apply directly here: https://forms.gle/WAMtYRujLubF8S6f9

V. INITIAL DATASET POPULATION: RAPID REVIEW

We conducted a rapid review of the literature, following the Cochrane guidelines [24] and PRISMA approach to reporting [22], with modifications to account for non-medical literatures[5]. Rapid reviews are "a type of knowledge synthesis in which [systematic review] methods are streamlined and processes are accelerated to complete the review more quickly" [24, p. 14]. Four databases representing general and engineering topics were searched: Scopus, Web of Science, IEEE Xplore, and ACM Digital Library. Searches were conducted on September 15th and 16th, 2021. This protocol was registered on OSF[6] before queries were run on September 10th, 2021.

From an initial set of 551 records, we screened the abstracts of 512, screened the full text of 44, and arrived at a final set of 18 articles reporting on a total of 19 experiments. All records at each stage of the process are available in the dataset. In effect, the rapid review process generated the initial records for the dataset as well as provided a first test of its structure with data from a representative sample of real articles.

VI. CONCLUSION

The field of HRI is maturing even as hot new topics emerge. As a form of science, HRI research can, and arguably, should, adapt to the trend and standards of science as a discipline. Open science methodologies and systematic review work, in particular living reviews that provide an open access dataset, represent the latest initiatives and scholarly rigour. We have chosen to focus on gender neutrality in robots, a challenging and exciting field of study within HRI that is growing in momentum. We have offered a framework and methodology, including a formalized submission tool, for conducting a living review going forward. We have also provided an open access database that has been validated with data from a rapid review, representing the first cycle of the living review process. We hope to engage the HRI community, especially those interested in gender and gender neutrality, to further test and refine this approach, adapt it to other topics, and contribute research for building consensus.

*A. Limitations and Future Work*

Two researchers determined what data should be extracted from the articles for the dataset, i.e., what relevant features of the research for the topic of gender neutrality in robots should make up the structure of the dataset. Future work will involve seeking input from other researchers, such as through a conference workshop or panel. Finally, there is a bug on OSF for spreadsheet files: it can only support single headers. We are currently working with OSF to enable multiple headers.

---

[4] https://sites.google.com/view/ro-man21-genr-workshop/home

[5] For example, we did not use a structured abstract or PICOS.

[6] https://osf.io/gcesd